\documentclass[12pt]{article}

\usepackage{epsfig}
\usepackage{epsf}

\def \ms {{\overline{\mbox{MS}}}}

\newcommand{\bea}{\begin{eqnarray}}
\newcommand{\eea}{\end{eqnarray}}

\begin{document}

\begin{center}
{\bfseries 
SMALL-X BEHAVIOR OF PARTON DENSITIES AT LOW Q2 VALUES
}

\vskip 5mm

A.Yu. Illarionov$^{1 \dag}$ and A.V. Kotikov$^{2 \dag}$ 

\vskip 5mm

{\small
(1) {\it
SISSA
and INFN, 
Sezione di Trieste, Trieste, Italy
}
\\
(2) {\it
BLThPh, JINR, 141980 Dubna (Moscow resion), Russia
}
\\
$\dag$ {\it
E-mail: illario@sissa.it, kotikov@theor.jinr.ru
}}
\end{center}

\vskip 5mm

\begin{center}
\begin{minipage}{150mm}
\centerline{\bf Abstract}
In the
leading twist approximation of the Wilson operator product expansion
with standard and ``frozen''
versions
of strong copling constant we show that
the Bessel-inspired
behavior 
of the structure function $F_2$ at small $x$, 
obtained for
a flat initial condition in the DGLAP evolution, 
leads to very good agreement with the 
deep inelastic 
scattering  experimental data from HERA.
\end{minipage}
\end{center}

\vskip 10mm

\section{Introduction}

The measurements of the deep-inelastic scattering
structure function
(SF)
$F_2$ 
at HERA
\cite{H1}
have permitted the access to
a very interesting kinematical range for testing the theoretical
ideas on the behavior of quarks and gluons carrying a very low fraction
of proton momentum,
the so-called small $x$ region.
In this limit one expects that
the non-perturbative effects may give an essential contributions. However, the
reasonable agreement between the HERA data and the 
next-to-leading order (NLO)
approximation of
perturbative
QCD, which has been observed for $Q^2 > 1 $GeV$^2$ (see 
the 
reviews
in \cite{CoDeRo}), indicates that the
perturbative QCD could describe the SF
evolution 
up to very low $Q^2$ values,
traditionally explained by soft processes.
It is of fundamental importance to find out the kinematical region where
the well-established perturbative QCD formalism
can be safely applied at small $x$.

The standard program to study the small $x$ behavior of
quarks and gluons
is carried out by comparison of the data
with the numerical solution of the
DGLAP
equations 
fitting the parameters of the
$x$ profile of partons at some initial $Q_0^2$ and
the QCD energy scale $\Lambda$ (see, for instance, \cite{fits,KKPS}).
However, 
in analyzing 
exclusively the
small $x$ region ($x \leq 0.01$), 
there is the alternative of doing a simpler analysis
by using some of the existing analytical solutions of DGLAP 
in the small $x$ limit (see \cite{CoDeRo} for review).
It was done 
in Refs. 
\cite{BF1}-\cite{Q2evo3},
where it was pointed out that the HERA small $x$ data can be
interpreted in 
the so called doubled asymptotic scaling (DAS) approximation
related to the asymptotic 
behavior of the DGLAP evolution 
discovered  in \cite{Rujula}
many years ago. 

Here we illustrate results obtained 
in \cite{Q2evo}-\cite{Q2evo3}, which 
are the extension  of previous 
leading order (LO)
studies \cite{Rujula,BF1} to the NLO QCD approximation.
The main ingredients are:

{\bf 1.} Both, the gluon and quark singlet densities are
presented in terms of two components ($'+'$ and $'-'$)
which are obtained from the analytical $Q^2$
dependent expressions of the corresponding ($'+'$ and $'-'$)
parton distribution (PD) moments.

{\bf 2.} The $'-'$ component is constant
at small $x$, whereas the 
$'+'$ component grows at $Q^2 \geq Q^2_0$ as 
$\sim \exp{(\sigma_{NLO})}$, where
\bea
\sigma_{NLO} = 2\sqrt{(\hat d_{+}s+\hat D_{+}p)\ln(x)},
\nonumber
\eea
the LO term $\hat d_+ = -12/\beta_0$ and the NLO one 
$\hat D_{+}=\hat d_{++}+\hat d_{+}\beta_1/\beta_0$ with
$\hat d_{++} = 412f/(27\beta_0)$. 
Here the coupling constant
$a_s=\alpha_s/(4\pi)$, 
$s=ln[a_s (Q^2_0)/a_s (Q^2)]$ and
$p=a_s (Q^2_0) - a_s (Q^2)$,
$\beta_0$ and $\beta_1$ are the first two 
coefficients of QCD 
$\beta$-function and $f$ is the number of active flavors.

\section{Basic formulae
}

Our purpose
is to extract
the small $x$ asymptotic PD
form 
in the framework of the DGLAP equation starting at some $Q^2_0$ with
the flat function:
 \begin{eqnarray}
f^{\tau2}_a (Q^2_0) ~=~
A_a ~~~~(\mbox{ hereafter } a=q,g), \label{1}
 \end{eqnarray}
where $f^{\tau2}_a$ are the leading-twist parts of
parton (quark and gluon)
distributions, multiplied by $x$,
and $A_a$ are unknown parameters that have to be determined from data.
We neglect
the non-singlet quark component at small $x$.

We would like to note that HERA data \cite{H1} show a rise
of $F_2$ 
at low $Q^2$ values ($Q^2 < 1 $GeV$^2$)
when $x \to 0$ (see  Fig.1 
below). This rise can be explained naturally
by incorporation  of higher-twist terms in the
analysis (see \cite{Q2evo3} and  Fig.1).

We shortly compile below the LO results 
(the NLO 
results 
may be found 
in \cite{Q2evo,Q2evo3}). 
%
The full small $x$ asymptotic results
for PD
 and SF $F_2$ 
at LO 
are:
 \begin{eqnarray}
F_2(x,Q^2)&=& e \, f_q(x,Q^2),
\label{r10} \\ 
 f_a(x,Q^2) &=& f^{+}_a(x,Q^2) + 
f^{-}_a(x,Q^2) \; , 
\label{r11}
\end{eqnarray}
where $e=(\sum_1^f e_i^2)/f$ is the average charge square.
The $'+'$ and $'-'$ components $f^{\pm}_a(x,Q^2)$
are given by the sum
 \begin{eqnarray}
 f^{\pm}_a(x,Q^2) ~=~ f^{\tau2,\pm}_a(x,Q^2) + 
f^{h\tau,\pm}_a(x,Q^2) \;  
\label{r12}
\end{eqnarray}
of
the leading-twist parts $f^{\tau2,\pm}_a(x,Q^2)$ 
and the 
higher-twist parts $f^{h\tau,\pm}_a(x,Q^2)$, 
respectively.

The small $x$ asymptotic results for PD $f^{\tau2,\pm}_a$ are 
%
%
\begin{eqnarray}
f^{\tau2,+}_g(x,Q^2)&=& \biggl(A_g + \frac{4}{9} A_q \biggl)
\tilde I_0(\sigma) \; e^{-\overline d_{+}(1) s} ~+~O(\rho) 
~~\;\; ,\label{8.0} \\
f^{\tau2,+}_q(x,Q^2)&=& \frac{f}{9}\biggl(A_g + \frac{4}{9} A_q \biggl) 
\rho \; \tilde I_1(\sigma) \;
e^{-\overline d_{+}(1) s} ~+~O(\rho) \; , \label{8.01} 
\\
f^{\tau2,-}_g(x,Q^2)&=& - \frac{4}{9} A_q e^{- d_{-}(1) s} 
~+~O(x) ,
\label{8.00} \\
f^{\tau2,-}_q(x,Q^2)&=&  A_q e^{- d_{-}(1) s} ~+~O(x) \; ,\label{8.02}
 \end{eqnarray}
where
$\overline d_{+}(1) = 1+20f/(27\beta_0)$ and
$          d_{-}(1) = 16f/(27\beta_0)$
are the regular parts of $d_{+}$ and $d_{-}$
anomalous dimensions, respectively, in the limit $n\to1$ 
\footnote{
For a quantity $k(n)$ we use the notation
$\hat k(n)$ for the singular part when $n\to1$ and
$\overline k(n)$ for the corresponding regular part. }. 
%
%
The functions $\tilde I_{\nu}$ ($\nu=0,1$) 
are related to the modified Bessel
function $I_{\nu}$
and to the Bessel function $J_{\nu}$ by:
\begin{eqnarray}
\tilde I_{\nu}(\sigma) ~=~
\left\{
\begin{array}{rl}
I_{\nu}(\sigma), & \mbox{ if } s \geq 0 \\
J_{\nu}(\sigma), & \mbox{ if } s  <   0
\end{array} \right.  
.
\label{4}
\end{eqnarray}
The 
variables $\sigma$ and $\rho$ are
given by
\begin{eqnarray}
\sigma =2\sqrt{\hat d_{+} s \ln(x)} \; , ~~~
\rho = \sqrt{\frac{\hat d_{+} s}{\ln(x)}}
= \frac{\sigma}{2\ln(1/x)}
\label{slo}
\end{eqnarray}
%
%


\section{Effective slopes} \indent


%
As it has been 
shown in Refs.\cite{Q2evo}-\cite{Q2evo3},
the PD and $F_2$ behavior, 
given in the Bessel-like form 
by generalized DAS approach,
can mimic a power law shape
over a limited region of $x$ and $Q^2$:
 \begin{eqnarray}
f_a(x,Q^2) \sim x^{-\lambda^{eff}_a(x,Q^2)}
 ~\mbox{ and }~
F_2(x,Q^2) \sim x^{-\lambda^{eff}_{F2}(x,Q^2)}.
\nonumber    \end{eqnarray}

At the twist-two LO approximation, they
have the following form 
 \begin{eqnarray}
\lambda^{eff}_g(x,Q^2) &=& \frac{f^+_g(x,Q^2)}{f_g(x,Q^2)} \,
\rho \, \frac{\tilde I_1(\sigma)}{\tilde I_0(\sigma)},
\nonumber
\\
\lambda^{eff}_{F2}(x,Q^2) &=&
\lambda^{eff}_q(x,Q^2) ~=~ \frac{f^+_q(x,Q^2)}{f_q(x,Q^2)} \,
\rho \, \frac{\tilde I_2(\sigma)}{\tilde I_2(\sigma)}.
\label{10.1}
\end{eqnarray}
%
The corresponding NLO expressions and the 
higher-twist terms
can be found in Refs. \cite{Q2evo}-\cite{Q2evo3}.

The effective slopes $\lambda^{eff}_a $ and 
$\lambda^{eff}_{F2}$ depend on the magnitudes $A_a$ of the initial PD
and also on the chosen input values of $Q^2_0$ and $\Lambda $.
To compare with the experimental data it is necessary to use 
the exact expressions
(\ref{10.1}) but for qualitative analysis one can
use some appropriate approximations.

At 
large values of $Q^2$, 
the ``$-$'' component of PD is negligable and
the dependence of slopes on the 
PD disappears.
In this case 
the asymptotic behaviors of slopes have the following expressions
\footnote{The asymptotic formulae given in Eq. (\ref{11.1})
work quite well at any $Q^2 \geq Q^2_0$ values,
because at $Q^2=Q^2_0$ the values of
$\lambda^{eff}_a $ and $\lambda^{eff}_{F2}$ are equal zero. 
The use of approximations in Eq. (\ref{11.1}) instead of the exact results 
given in Eq. (\ref{10.1}) underestimates 
(overestimates) only slightly the gluon (quark) slope
at $Q^2 \geq Q^2_0$.
}:
 \begin{eqnarray}
\lambda^{eff,as}_g(x,Q^2) &=& 
\rho \, \frac{\tilde I_1(\sigma)}{\tilde I_0(\sigma)} \approx \rho - 
\frac{1}{4\ln{(1/x)}} ,
\nonumber \\
\lambda^{eff,as}_{F2}(x,Q^2) 
&=&
\lambda^{eff,as}_q(x,Q^2) ~=~ 
\rho \, \frac{\tilde I_2(\sigma)}{\tilde I_1(\sigma)} \approx \rho - 
\frac{3}{4\ln{(1/x)}}, 
\label{11.1} 
\end{eqnarray}
where the symbol $\approx $ marks the approximation obtained by the  expansion
of the usual and modified Bessel functions in (\ref{4}).

One can observe from (\ref{11.1}),  that
the gluon effective slope $\lambda^{eff,as}_g$ 
is larger than the quark slope
$\lambda^{eff,as}_q$, which is in excellent agreement with  global analyses 
\cite{fits}.

\section{Comparison with experimental data
} \indent

With the help of the results presented in the previous sections we have
analyzed $F_2$ HERA data at small $x$ from the H1 and ZEUS collaborations
\cite{H1}.
In order to keep the analysis as simple as possible
we have fixed the number of active flavors $f$=4 and
$\Lambda_{\ms}(n_f=4) = 292$ MeV in agreement
with the more recent H1 results \cite{H1slo}.

The typical fits for the SF $F_2(x, Q^2)$ as a function of $x$ for
different $Q^2$ bins are presented in Fig. 1.
The experimental points are from H1 
(open points) 
and ZEUS 
(solid points).
The solid line represents the NLO fit alone with
$\chi^2/{n.d.f.} = 1.31$.
The dashed curve are obtained from the fit at the NLO, when the
renormalon contributions of higher-twist terms have been incorporated.
The corresponding
$\chi^2/{n.d.f.} = 0.86$.
The dash-dotted curve (hardly distinguished from the dashed one) represents
the fit at the LO together with the renormalon contributions of higher-twist
terms. The corresponding
$\chi^2/{n.d.f.} = 0.84$.
The results demonstrate excellent agreement between theoretical predictions
and experimental data for the region $Q^2 \ge 0.5$ GeV$^2$.
However, the twist-two appriximation is in agreement with the data only for
$Q^2 \ge 2.5$ GeV$^2$.

Using these results of the fits of the SF $F_2(x,Q^2)$
we analyze also the HERA data for the slope $d\ln F_2/d\ln (1/x)$
at small $x$ from the H1 and ZEUS Collaborations 
\cite{H1slo}-\cite{DIS02}.
 The results are shown in Fig. 2.
Because the twist-two
approximation is reasonable at $Q^2 \geq 2.5$ GeV$^2$ only,
some modification 
should be considered at the lower $Q^2$ values.
In the paper
\cite{Q2evo3} we have added the higher twist corrections and have found a good
agreement for 
$Q^2 \geq 0.5$ GeV$^2$.

Here we study another possibility. We modify the QCD coupling constant. From 
different studies \cite{DoShi,bfklp} it is known that the effective
argument of the coupling constant is higher then $Q^2$ at low $x$
values.
In the present paper we consider freezing
of the coupling constant by changing its argument 
$Q^2 \to Q^2 + M^2_{\rho}$,
where $M_{\rho}$ is the $\rho $-meson mass \cite{Greco}. Thus, in the 
formulae of the
Section 2, we should do the following replacement
\begin{equation}
 a_s(Q^2) \to a_{fr}(Q^2) \equiv a_s(Q^2 + M^2_{\rho})
\label{Intro:2}
\end{equation}


Fig. 2 shows the experimental data for $\lambda_{F2}$.
It demonstrates an essential improvement between
the low $Q^2$ ZEUS data for $\lambda^{eff}_{F2}(x,Q^2)$ and our
results with the ``frozen'' coupling constant $a_{fr}(Q^2)$.

Indeed, from our fits for $F_2(x,Q^2)$ we have found that $Q^2_0 \approx
0.5 \div 0.8$ GeV$^2$ \cite{Q2evo3}.
 So, initially we had $\lambda^{eff}_{F2}(x,Q^2_0)=0$
from the suggestion (\ref{1}). The replacement (\ref{Intro:2}) modifies the
value of $\lambda^{eff}_{F2}(x,Q^2_0)$. 
For the  
``frozen'' coupling constant $a_{fr}(Q^2)$ the value of
$\lambda^{eff}_{F2}(x,Q^2_0)$ is non-zero 
and the slope 
is quite close to experimental data at $Q^2 \approx 0.5$ GeV$^2$.
Nevertheless, for  $Q^2 \leq 0.5$ GeV$^2$ there is still an disagreement with
the data, that needs additional investigations.

\begin{figure}[t]
\vskip -0.5cm
\epsfig{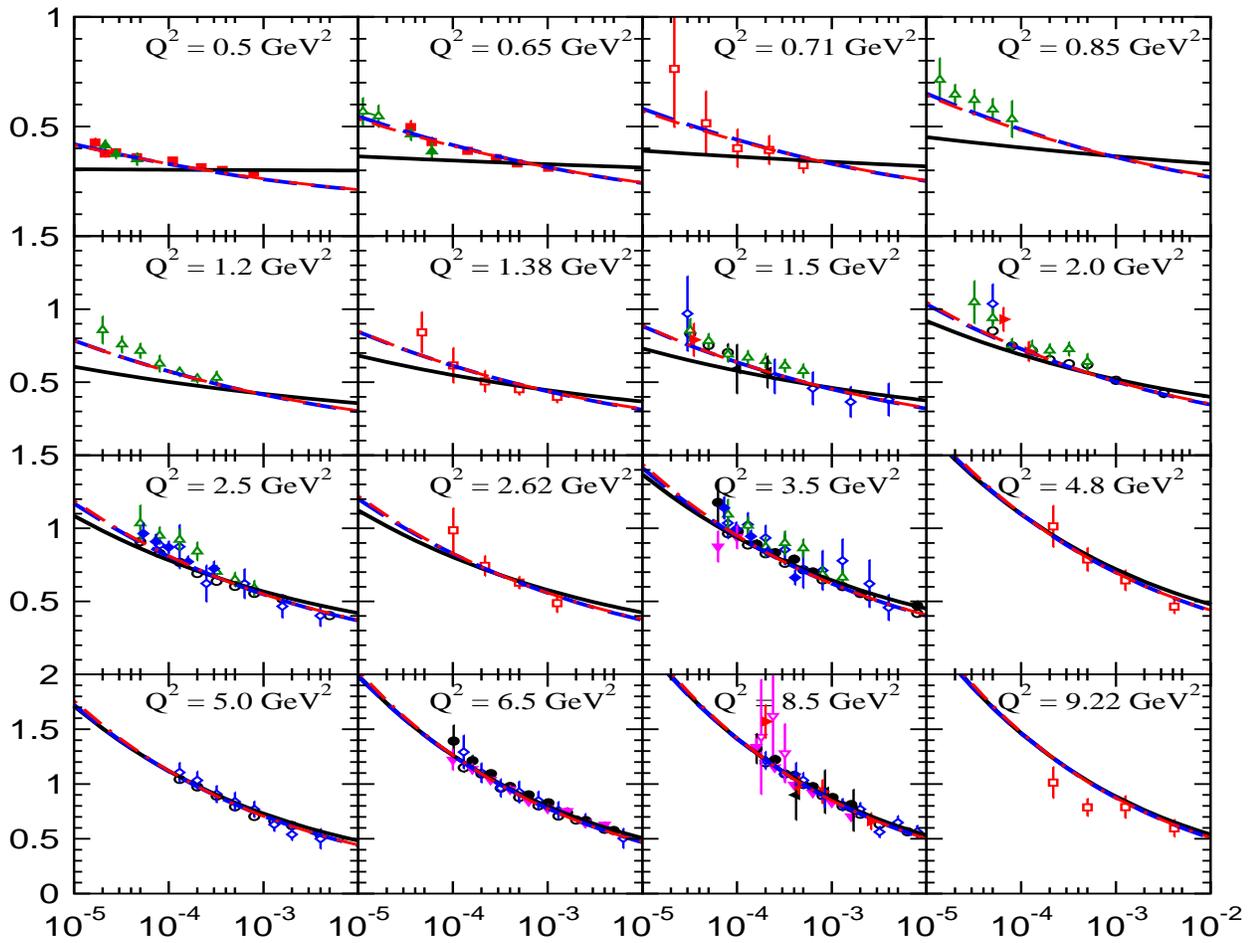}
%
\caption{The structure function $F_2$ as a function of $x$ for different
$Q^2$ bins. 
}
\end{figure}

\begin{figure}[t]
\vskip -0.5cm
\psfig{figure=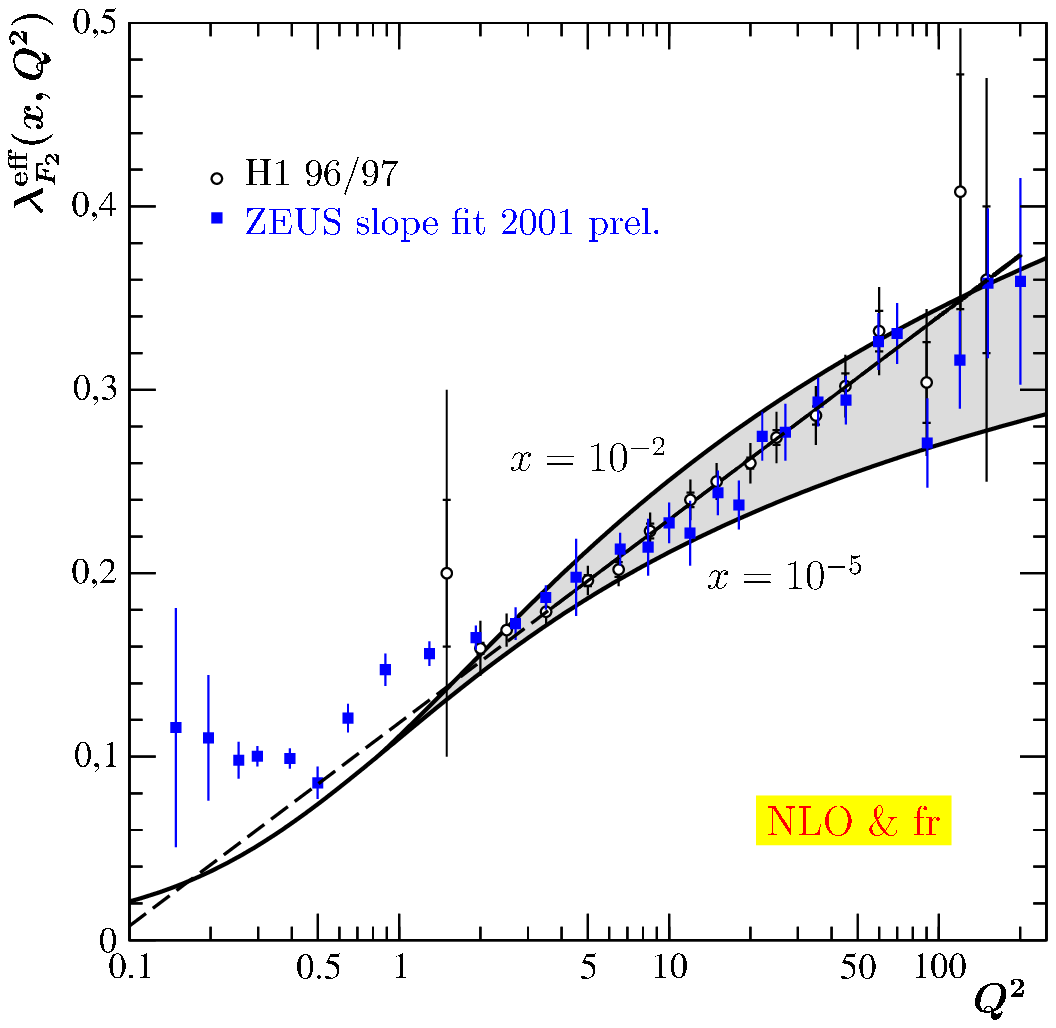,height=4.9in,width=6.5in}
%
\caption{The slope $\lambda^{eff}_{F2}(x,Q^2)$
as a function of 
$Q^2$.
}
\end{figure}

\section{Conclusions} 

We have shown that the results, 
developed recently in \cite{Q2evo}-\cite{Q2evo3},
 have quite simple form and reproduce many PD
properties 
at small $x$,
that have been known from global fits.

We found the very good agreement between our approach, based on QCD, 
and HERA data, as it has been observed earlier with
other approaches (see reviews \cite{CoDeRo}). 
%
The (renormalon-type) higher-twist terms lead to the natural explanation of
the rise of SF $F_2$ 
at low values of $Q^2$ and $x$, which
has been discovered in recent HERA experiments \cite{H1}.

The application of the ``frozen'' coupling constant $a_{fr}(Q^2)$
leads to good agreement with the recent HERA data  \cite{H1slo}-\cite{DIS02}
for the slope $\lambda^{eff}_{F2}(x,Q^2)$ for  $Q^2 \geq 0.5$ GeV$^2$.
As the next step of our investigations, we plan to fit the HERA experimental
data for the $F_2(x,Q^2)$ SF directly with the ``frozen'' coupling constant
$a_{fr}(Q^2)$.

  {\it Acknowledgments.}
One of the authors (A.V.K.)
 would like to express his sincerely thanks to the Organizing
 Committee 
for the kind invitation.
He was supported in part, 
by Heiserberg-Landau program
and by the Russian Foundation for Basic Research (Grant N 08-02-00896-a).

\end{document}